# Can quantum physics help solve the hard problem of consciousness? A hypothesis based on entangled spins and photons

*Christoph Simon, Department of Physics and Astronomy & Institute for Quantum Science and Technology, University of Calgary, Calgary, AB T2N1N4, Canada. Email: csimo@ucalgary.ca*

**Introduction**

There are still many open questions in the sciences of the brain and mind (Adolphs 2015), but the deepest one, and the one where we really haven't made much progress, is what Chalmers (1995) called the hard problem of consciousness: how does subjective experience arise from brain matter? This question has occupied philosophers and scientists for centuries. Arguably the recent rapid development of artificial intelligence adds a new urgency (Tegmark 2017). We might soon have machines that seem like conscious beings in many respects. It would seem important to understand if they really are, keeping in mind also that arguments by analogy are less convincing in the case of machines than they are for animals. In the following I suggest exploring the possibility that quantum physics may have something to offer in this respect. Building on recent progress in quantum technology and neuroscience, I propose a concrete hypothesis as a basis for further investigations, namely that subjective experience is related to the dynamics of a complex entangled state of spins, which is continuously generated and updated through the exchange of photons. I begin by describing my main arguments in favor of quantum physics playing a role in subjective experience. They address the two most common questions/objections that are raised in response to the 'quantum consciousness' thesis, namely (first) why do we need quantum physics, and (second) could it possibly work? Then I outline several additional arguments, before turning to the question of physical and biological plausibility and explaining the motivations for the proposed concrete hypothesis involving spins and photons. I point out various opportunities for experimental tests across a range of scientific disciplines. Finally I discuss a number of philosophical questions raised by the present proposal.

**Main arguments**

Why might we need quantum physics to solve the hard problem of consciousness? Quantum entanglement offers a unique solution for the binding problem of conscious experience (Marshall 1989). Conscious experience is *complex, yet unified*. However, systems described by classical physics can always be analyzed in terms of the dynamics of their parts, without sacrificing anything but convenience. This makes the unity of consciousness - which is definitely complex - very mysterious from a classical perspective. In quantum physics the situation is fundamentally different (Dhand *et al.* 2018, Raymer 2017). A quantum system that is entangled is more than the sum of its parts in a strong sense that is well-defined physically and mathematically (Horodecki *et al.* 2009). Quantum entanglement has a fundamentally holistic character. For an entangled system, the state of the whole is completely well-defined, while the states of subsystems are indeterminate. There is no analogy for this phenomenon in classical physics[1]. Entanglement is the foundation for quantum non-locality, i.e. the fact that the behavior of entangled particles can be strongly correlated even when they are separated by long distances. Bell's theorem (Bell 1964) proves mathematically that entanglement is fundamentally different from classical physics concepts. Like subjective experience, entangled quantum systems can be unified and complex at the same time. This provides strong motivation to explore the possibility that subjective experience might be related to the existence of quantum entanglement in the brain.

But could it really work? Won't fragile quantum effects like entanglement be quickly destroyed in the wet, warm environment of the brain? (Tegmark 2000) This is where things have changed significantly in recent years. Quantum scientists and engineers are now developing quantum technology that is sophisticated and complex, including condensed-matter systems where such quantum phenomena can be observed at room (or body) temperature. We have learned a lot about which physical systems and degrees of freedom are promising to use in this context, in particular spins and photons. I will describe this recent progress in more detail and develop a concrete hypothesis that uses these elements below.

**Additional arguments**

There are several additional notable arguments that can be made and have been made for quantum physics playing a role in the brain. One line of thought is inspired by the special role of the observer in quantum physics. It is well known that in the conventional ('Copenhagen') formulation of quantum physics the act of observation plays a key part in the dynamics - it collapses the wave function, whose evolution is otherwise described by the Schrodinger equation. Maybe this special status of the observer means that there is a close connection between consciousness and quantum physics (Stapp 2009)? For example, maybe the wave function collapse is due to the observer's subjective experience (Wigner 1967)? A different line of arguments involving both consciousness and quantum wave function collapse was developed by Penrose (Hameroff and Penrose 2014). He argued that mathematical insight is non-computational, and that the non-computational element might be provided by a wave function collapse that is furthermore related to the unification of quantum physics and gravity. It is possible to formulate quantum physics in a way such that wave function collapse is avoided, this is known as the 'many-worlds' interpretation (Everett 1957). But even in this case one has to think about the quantum states corresponding to the observer's subjective experience (Lockwood 1989, Simon 2009). These arguments show at the very least that the topic of subjective experience is difficult to completely avoid in quantum physics.

---

[1] I would therefore argue that the type of unification offered by quantum entanglement is much more fundamental than that offered by classical concepts such as, for example, integrated information (Tononi *et al.* 2016).

A somewhat related point is that the concept of matter is a much more subtle concept in quantum physics than in classical physics. The existence of quantum entanglement and the seemingly special role of the observer are all aspects of the fact that quantum systems – which are after all the basic constituents of matter - behave in highly non-intuitive, seemingly 'non-material' ways. This in itself may give one hope that quantum physics can help in resolving the mind-matter problem.

Finally a recent line of argument is inspired by the development of quantum computing, which aims to use the unique features of quantum physics to solve problems that are difficult for conventional computers (Nielsen and Chuang 2000; Raymer 2017). Maybe nature discovered quantum computing before we did and implemented it in the brain (Fisher 2015)? Note that following this line of argument quantum effects could in principle have a purely 'functional' role, not necessarily related to subjective experience.

**Physical and biological plausibility**

Let me turn to the question of physical and biological plausibility. Quantum technology as a whole is relatively recent, but quantum computers are now being developed in earnest, including by industrial actors such as IBM and Google. Important concepts like quantum error correction and fault tolerant quantum computing have been developed (Nielsen and Chuang 2000). These developments show that a quantum system can in principle be kept in a complex entangled state for long times while performing useful computational tasks. Equally importantly for the present discussion, until very recently most quantum technology required very low temperatures or ultra-high vacuum conditions. The brain therefore seemed like a highly unlikely environment for observing fragile quantum phenomena. The initial focus in these discussions was on electrical phenomena (Hameroff and Penrose 2014, Tegmark 2000), maybe because the brain definitely uses electrical signals for classical signaling. However, modern quantum technology suggests a different path. Electrical phenomena can be good for observing quantum effects at low temperatures or in vacuum. Superconducting quantum circuits and trapped ions are good examples (Ladd *et al.* 2010). But if the goal is to observe quantum phenomena at room (or body) temperature in condensed-matter systems, then spins and photons are much more promising. The reason is that spins and photons interact much more weakly with their environment than electrical charges do, and the interaction with the environment is what typically destroys quantum phenomena, in a process known as environmentally induced decoherence (Schlosshauer 2007).

Nuclear spins in small molecules in liquid solution at room temperature were in fact used early on for demonstrations of quantum information processing (Ladd *et al*. 2010). This approach was not scalable, but it showed that nuclear spins can exhibit subtle quantum properties even under these seemingly challenging conditions, and that they can do so on the millisecond to second timescales that are relevant for subjective experience. More scalable approaches to quantum information processing using electron and nuclear spins in solids at room temperature are now being developed (Dolde *et al*. 2013; Maurer *et al*. 2012). To my knowledge, the possibility of quantum effects related to spins playing a role in the generation of consciousness was first explored in detail by Hu and Wu (2004). More recently Fisher (2015) has developed a proposal how quantum computing might happen in the brain based on nuclear spins.

Photons are the system of choice in quantum technology whenever the goal is to transmit quantum information or entanglement over macroscopic distances. The development of quantum networks – ultimately extending to global distances – is currently a very active area of research (Simon 2017). The current distance record is held by the Chinese quantum communication satellite Micius, which established entanglement over more than a thousand kilometers (Yin *et al.* 2017). Such quantum networks will allow applications such as distributed quantum computing, secure communication, and quantum sensing. A quantum network typically requires photons for communication, but also stationary quantum nodes – such as spins – that can store and process the quantum information. Single-photon sources and single-photon detectors are also important components. Researchers including my own group are working on developing key components for quantum networks that can operate at room temperature, such as sources of single photons with the right quantum-physical properties (Wein *et al.* 2018) and interfaces between photons and spins (Ghobadi *et al.* 2017).

Given this context, it is very interesting that neurons can emit photons (Kobayashi *et al.* 1999, Tang and Dai 2014). Reactive oxygen species generated in the mitochondria are thought to be one likely source. While the emission rates are very low - maybe of the order of one photon per neuron per minute (Kumar *et al.* 2016)  -, this also means that they are naturally at the quantum level, which could be viewed as advantageous in the current context. Could these photons be used for communication? My focus here is on potential quantum communication, but even ordinary (classical) cellular communication via light would be a significant discovery, and this possibility has been investigated for quite some time (Chang et al. 1998). One argument in favor of photonic cellular communication is the existence of light-sensitive proteins – opsins - throughout the body. These opsins are evolutionarily conserved, which suggests that they serve a biological function (Popejoy 2017). So both photon sources and single-photon detectors seem to exist in the body, and in particular in the brain. There is also some evidence that mammalian cerebral tissue responds to low-level light (Wade *et al.* 1988). Let me note that a role of photons in cellular signaling is made more plausible by the fact that they are linked to reactive oxygen species, which are very dangerous for cells. This means that it could have had great evolutionary value for the cells to learn to detect these photons, and later on this evolved ability could have been used for more general communication purposes. A similar explanation has been proposed for the evolution of action potentials out of an original cellular response to mechanical threats to the cell membrane (Brunet and Arendt 2016).

The use of photons for communication in the brain would require a way for the photons to target specific cells. This is important for the communication to be selective, and to avoid unwanted effects from ambient light. It is well known that axons serve as targeted conductors of electrical signals between neurons. With a group of collaborators I recently suggested that they could serve as conductors for photons (photonic waveguides) as well (Kumar *et al.* 2016). The key property in this context is the refractive index, which needs to be higher than that of the surrounding medium for light guidance to be possible. It turns out that this is indeed the case for axons. If the axons are wrapped with a myelin sheath, as they often are, the refractive index is higher still. In this context it is interesting to mention the Muller cells in the retina, which are glia cells, just like the cells that form the myelin sheaths around axons. It was recently shown experimentally that Muller cells serve as light guides in the retina (Franze *et al.* 2007).

Despite this encouraging evidence, one might still have doubts about light guidance by axons. Unlike man-made optical fibers, axons are not perfectly smooth and regular, but have many imperfections from an optical perspective. Can they still be sufficiently good waveguides to allow communication between different neurons in the brain? We addressed this question via detailed numerical modeling (Kumar *et al.* 2016). The biggest concern was the existence of Ranvier nodes, where the myelin sheath around the axon is interrupted. We constructed a detailed numerical model for these nodes using techniques from photonics research. We concluded that the transmission through these nodes can be high, depending on the exact dimensions, which have significant natural variation. We also studied many other 'imperfections', including variations in cross-sectional area, non-circular cross-sections, and bends. In all these cases we concluded that reasonable light transmission is feasible for biologically realistic parameters. We also proposed a number of potential experimental tests. A very direct test would be to isolate living neurons and couple light into their axons to test the light guidance. Other approaches involve introducing artificial sources (such as nanoparticles) and detectors (such as rhodopsins) in vivo. Pinpointing the sources of the photons in the cells more precisely would also be of great interest. On the basis of our modeling results, we estimated the achievable range and rates for the hypothetical photonic communication. We predict transmission probabilities ranging from a few percent to almost 100% for the relevant distances in the millimeter to centimeter range, at least for certain axons. Given that of the order of a billion photons seem to be emitted per second throughout the brain (Tang and Dai 2014, Kumar *et al.* 2016), there appears to be the potential for a significant amount of information or entanglement to be transmitted.

The concrete hypothesis that I propose as a basis for further investigations is that subjective experience is related to the dynamics of a complex entangled state of specific spins, which is continuously generated and updated through the exchange of photons. The precise form of this entangled state would of course have to be influenced by a lot of classical information coming from the senses and from the brain itself. The special role of the photons, or of some of the photons, would be to maintain the quantum entanglement, a task for which they are uniquely suited, as I have discussed above when talking about quantum networks. This leads to the question of how many photons would need to be exchanged per second to make such a scenario consistent with what we know about subjective experience. There does not seem to be an established consensus regarding the 'bandwidth' of subjective experience, but some upper and lower bounds can be derived from the literature. Both brain architecture and subjective experience suggest that our bandwidth is probably dominated by visual experience. For a lower bound, psychophysics experiments on reading comprehension show that the bandwidth is at least of the order of 100 bits per second (Zimmermann 1989). For an upper bound, theoretical work on modeling visual information processing in the eye and brain suggests that our subjective visual bandwidth is probably not greater than 100,000 bits per second (Watson 1987, Rojer and Schwartz 1990), but possibly much smaller. It would be very interesting to design and conduct psychophysics experiments that could give a more precise value for this bandwidth. But the present bounds already seem to imply that a rate of a billion photons per second is in principle more than sufficient.

Where exactly might this entangled state be located in the brain? The question of the seat of subjective experience is a topic of lively debate in neuroscience, independently of the present quantum physical considerations. There are very different proposals, ranging from theories according to which the key systems involved in subjective experience would be located in the cortex (Koch *et al.* 2016) to theories where specific parts of the midbrain are the main generators of subjective experience, whereas the cortex plays only a supporting role (Damasio 2010, Merker 2007), with intermediate theories focusing on the role of the thalamus (Ward 2011). The proposed hypothesis would be somewhat more compatible with smaller structures being the generators of consciousness, mostly because it is easier for photons to be transmitted in axons over millimeter compared to centimeter distances, but the relevant cortical distances may also not be out of reach. It is worth noting that the question of the seat of subjective experience is closely related to the question of its evolutionary age. For example, if crabs (Elwood and Appel 2009) or insects (Barron and Klein 2016) have subjective experience, then it is more likely to be localized in more basic brain structures such as the midbrain. The common ancestor of crabs and humans, which might already have had some level of subjective experience in this case, was probably quite a simple worm-like bilaterian. The bandwidth of subjective experience would presumably be quite small in such simple animals.

Let me close this section by briefly discussing some open physics questions. I have argued that spins and photons might both play key roles, the former because they can keep their quantum properties intact over the relatively long timescales that are relevant for subjective experience (milliseconds to seconds), the latter because they are ideal for distributing quantum entanglement over the macroscopic distances that would presumably be required to generate entangled states of sufficient complexity. How would the spins and the photons interface with each other? Spin-photon interfaces already exist in quantum network research (Simon 2017), including proposals for operation at room temperature (Ghobadi *et al.* 2017). Imagining a solution that could work in the brain is an interesting challenge. One potentially promising system is the oxygen molecule, which already played an important role in the proposal of (Hu and Wu 2004) because of its non-zero electronic spin and its abundance in the body. In that proposal the electronic spins of small paramagnetic molecules (oxygen or nitrous oxide) are thought to modulate the entanglement of nuclear spins in the cell membranes. In the present context oxygen is of interest not only because of its spin, but also because reactive oxygen species, in particular singlet oxygen (Ogilby 2010), an excited state of the oxygen molecule, are likely candidates to be sources of the photons emitted by neurons (Kobayashi *et al.* 1999). Moreover the electronic spin of the oxygen molecule could conceivably act as a mediator between the photons emitted by the same molecule and nearby nuclear spins, which may be better candidates for long-time quantum entanglement than the electronic spin itself, because they interact more weakly with their environment. It is interesting to note that the radical pair theory of magnetoreception, which is one of the leading theories for how birds sense magnetic fields, is based on interactions between the same basic ingredients that I am invoking here – photons, electronic spins and nuclear spins (Hiscock *et al.* 2016). The present considerations have the potential to stimulate interesting quantum experiments with photons and molecular spins in liquids.

One of the challenges for spin-based explanations of cognition and subjective experience is the fact that high magnetic fields, e.g. in MRI scanners, do not disrupt consciousness, even though they change the energy states of the spins. The proposal of Fisher (2015) avoids this problem by assuming that the quantum information is represented by multiple singlet (i.e. total spin zero) states of several nuclear spins in a single molecule. Such singlet states are almost completely insensitive to external magnetic fields of the type experienced in MRI scanners. Similar ideas should work as well in the context of the present line of thought. In Fisher's proposal quantum entanglement is distributed between neurons through molecular diffusion, which is very slow. The main advantage of photons is that they would provide a much more convenient way of transmitting the entanglement, as is routinely done in man-made quantum technology (Simon 2017).

**Philosophical questions**

I would finally like to mention a few philosophical questions that are raised by the present hypothesis. In particular, how does it position itself relative to the concepts of materialism, panpsychism, dualism, and epiphenomenalism? Subjective experience could be related to quantum entanglement, but still be a fundamental feature of the universe, so the present hypothesis does not necessarily imply a materialist point of view. Does it imply a form of panpsychism? It is worth noting that entanglement seems to provide an elegant solution for the 'combination problem' of panpsychism (Bruntrup and Jaskolla 2016) – if small constituents of the universe have their own small subjective experience (which is the basic idea of panpsychism), then how can more complex systems still have a unified experience? How are many small experiences combined into one? If subjective experience corresponds to an entangled state, then this happens naturally. When the constituents become entangled with each other, they lose their individual definite states and indeed become parts of a coherent whole. This point is very similar to the discussion of the binding problem at the beginning of this paper. However, this doesn't necessarily mean that the present hypothesis favors strong forms of panpsychism. In particular, a tempting strong hypothesis might be to assume that every entangled state corresponds to a system that has subjective experience. But this quickly leads to inconsistencies because superpositions of entangled quantum states can also be entangled, but superpositions have a probabilistic meaning in quantum physics. A more viable possibility is that there is a specific basis of states in quantum physical state space that correspond to definite subjective experiences (Lockwood 1989; Simon 2009), and that the states forming this basis are entangled states.

The very definition of quantum entanglement relies on the existence of subsystems. But what are the appropriate subsystems to choose?[2] I have proposed that the appropriate subsystems might be individual spins, but there are subtle differences between the different options for the spins from a physics perspective. To the best of our current physical knowledge, electrons are elementary particles without further structure. In contrast, atomic nuclei are made of protons and neutrons, and the protons and neutrons themselves are made of quarks and gluons. This might suggest that electrons are a more promising choice for the present hypothesis, especially if one thinks as subjective experience as a very fundamental phenomenon. However, nuclear spins are more weakly coupled to their environment and therefore tend to keep their quantum properties over longer timescales, making them somewhat more attractive from the point of view of physical plausibility. There may be room for different versions of the basic hypothesis that favor different physical systems, or that are more or less specific regarding the physical systems that are thought to be related with subjective experience. Hypotheses that are very specific - e.g. only entanglement between electron spins is relevant for subjective experience – might be easier to formulate in a consistent way. Such a theory could probably be classified as dualist rather than panpsychist.

Finally, where does the present hypothesis stand on the question whether subjective experience is epiphenomenal or causally efficient? The structural match argument – both quantum entanglement and subjective experience are unified and complex – is consistent with the idea that conscious experience could be the subjective side of certain kinds of entangled systems. It is difficult to understand why we have subjective experience if it doesn't have a causal role, given the general parsimoniousness of evolution (Kent 2016). Entanglement can definitely be useful, e.g. for information processing tasks, as illustrated by quantum computing. So evolution might have selected entanglement for its usefulness, and subjective experience might be the unavoidable consequence of that selection. Would this be a form of epiphenomalism?  There may be room for different points of view on this question, as well as on the others raised in this section, even if one entertains the basic hypothesis proposed here.

**Conclusions**

I have proposed to seriously explore the possibility that quantum entanglement might have something to do with subjective experience. I focused on the argument of structural similarity - both phenomena are simultaneously complex and unified, a combination that is difficult to understand from a classical physics perspective. I mentioned additional arguments, such as the subtlety of the concept of matter in quantum physics and the potential usefulness of quantum computing. The recent dramatic progress in quantum technology makes the present hypothesis more believable than it would have been a few decades ago. Based on this technological experience, I have argued that spins and photons are particularly promising systems - spins can keep their quantum properties for the relevant timescales, while photons can distribute entanglement over macroscopic distances. There are spins and photons in the brain, as well as potential photon detectors. Axons can plausibly act as photonic waveguides, and the oxygen molecule is one potential spin-photon interface. Experimental tests of many aspects of these ideas are imaginable, in areas as diverse as neuroscience, psychophysics and quantum physics, and are already under development in some cases. Finally the present ideas also raise interesting philosophical questions where a dialogue between quantum physicists and philosophers might be fruitful.

---

[2] This can be seen as a version of the 'grain' problem that has been raised in the context of integrated information theory (Moon and Pae 2018).

**Acknowledgements**

I thank D. Chalmers, M. Colicos, T. Craddock, J. Dowling, M. Fisher, I. Fuentes, J. Guck, S. Hameroff, S. Hastings-Simon, B. Heyne, C. Jaeger, D. Kaszlikowski, J. Kellner, S. Kumar, A. Lamas, H. Morch, A. Popejoy, R. Prix, J. Tuszynski, L. Ward, S. Wein, P. Zarkeshian, and T. Zhong for useful discussions and helpful suggestions. However, the point of view expressed here is mine, as are any factual mistakes.

**References**

Adolphs, R. (2015), 'The unsolved problems of neuroscience', *Trends Cogn. Sci.,* **19**, pp. 173-175.

Barron, A.B. and Klein, C. (2016), 'What insects can tell us about the origins of consciousness', *Proc. Natl. Acad. Sci. USA*, **113**, pp. 4900-4908.

Bell, J.S. (1964), 'On the Einstein-Podolsky-Rosen Paradox', *Physics*, **1**, pp. 195-199.

Brunet, T. and Arendt, D. (2016), 'From damage response to action potentials: early evolution of neural and contractile modules in stem eukaryotes', *Phil. Trans. R. Soc. B,* **371**, 20150043.

Bruntrup, G. and Jaskolla, L. (eds. 2016), *Panpsychism: Contemporary Perspectives* (Oxford: Oxford University Press)

Chalmers, D.J. (1995), 'Facing up to the problem of consciousness', *Journal of Consciousness Studies*, **2**, pp. 200-219.

Chang, J. et al. (eds. 1998), *Biophotons* (Alphen aan den Rijn: Kluwer Academic)

Damasio, A. (2010), *Self Comes to Mind: Constructing the Conscious Brain* (New York: Vintage)

Dhand, I. *et al.* (2018), 'Understanding quantum physics through simple experiments: from wave-particle duality to Bell's theorem', *arXiv*:1806.09958.

Dolde, F. et al. (2013), 'Room-temperature entanglement between single defect spins in diamond, *Nature Physics* **9**, pp. 139–143.

Elwood, R.W. and Appel, M. (2009), 'Pain experience in hermit crabs?', *Animal Behaviour*, **77**, pp. 1243-1246.

Everett, H. (1957), '"Relative State" Formulation of Quantum Mechanics', *Rev. Mod. Phys.*, **29**, pp. 454-462.

Franze, K. et al. (2007), 'Muller cells are living optical fibers in the vertebrate retina', *Proc. Natl. Acad. Sci. USA*, **104**, pp. 8287-8292.

Fisher, M. (2015), 'Quantum cognition: The possibility of processing with nuclear spins in the brain', *Ann. Phys.*, **61**, pp. 593-602.

Ghobadi, R. *et al.* (2017), 'Towards a Room-Temperature Spin-Photon Interface based on Nitrogen-Vacancy Centers and Optomechanics, *arXiv*:1711.02027.

Hameroff, S. and Penrose, R. (2014), 'Consciousness in the universe: A review of the 'Orch OR' theory', *Phys. Life Rev.,* **11**, pp. 39-78.

Hiscock, H.G. et al. (2016), 'The quantum needle of the avian magnetic compass', *Proc. Natl. Acad. Sci. USA*, **113**, 4634-4639.

Horodecki, R. et al. (2009), 'Quantum entanglement', *Rev. Mod. Phys.*, **81**, pp. 865-942.

Hu, H. and Wu, M. (2004), 'Spin-mediated consciousness theory: possible roles of neural membrane nuclear spin ensembles and paramagnetic oxygen', *Medical Hypotheses*, **63**, pp. 633-646.

Kent, A. (2016), 'Quanta and Qualia', *arXiv*:1608.04804.

Kobayashi, M. *et al.* (1999), '*In vivo* imaging of spontaneous ultraweak photon emission from a rat's brain correlated with cerebral energy metabolism and oxidative stress', *Neurosci. Res.* **34**, pp. 103–113

Koch, C. et al. (2016), 'Neural correlates of consciousness: progress and problems', *Nature Reviews Neuroscience*, **17**, pp. 307-321.

Kumar, S. *et al.* (2016), 'Possible existence of optical communication channels in the brain', *Scientific Reports*, **6**, 36508.

Ladd, T.D. *et al.* (2010), 'Quantum computers', *Nature*, **464**, pp. 45-53.

Lei, H. et al. (2003), '*In vivo* 31P magnetic resonance spectroscopy of human brain at 7 T: an initial experience', *Magn. Reson. Med.*, **49**, pp. 199-205.

Lockwood, M. (1989), *Mind, Brain and the Quantum: The Compound 'I'* (Oxford: Basil Blackwell).

Marshall, I.N. (1989), 'Consciousness and Bose-Einstein Condensates', *New Ideas in Psychology* **7**, pp. 73-83.

Maurer, P.C. *et al*. (2012), 'Room-temperature quantum bit memory exceeding one second', *Science,* **336**, pp. 1283-1287

Merker, B. (2007), 'Consciousness without a cerebral cortex: A challenge for neuroscience and medicine', *Behavioral and Brain Sciences*, **30**, pp. 63-81.

Moon, K. and Pae, H. (2018), 'Making Sense of Consciousness as Integrated Information: Evolution and Issues of IIT'*, arXiv*:1807.02103.

Nielsen, M. and Chuang, I. (2000), *Quantum Computation and Quantum Information* (Cambridge: Cambridge University Press).

Ogilby, P.R. (2010), 'Singlet oxygen: there is indeed something new under the sun', *Chem. Soc. Rev.*, **39**, pp. 3181-3209.

Popejoy, A.B. (2017), *Illumination at the Intersections of Genomics and Public Health: A Study of Opsins, SPurS, Cluster Machine, and Ancestry* (PhD thesis, University of Washington).

Raymer, M.G. (2017), *Quantum Physics: What Everyone Needs To Know* (Oxford: Oxford University Press).

Rojer, A.S. and Schwartz, E.L. (1990), 'Design characteristics for a space-variant visual sensor with complex-logarithmic geometry'*,* Proceedings 10th Int. Conf. Pattern Recognition (New York: IEEE), DOI: 10.1109/ICPR.1990.119370

Schlosshauer, M.A. (2007), *Decoherence and the quantum-to-classical transition* (Berlin: Springer).

Simon, C. (2017), 'Towards a global quantum network', *Nature Photonics*, **11**, pp. 678-81.

Simon, C. (2009), 'Conscious observers clarify many worlds', *arXiv*:0908.0322

Stapp, H.P. (2009), *Mind, Matter and Quantum Mechanics* (Berlin: Springer).

Tang, R. and Dai, J. (2014), 'Spatiotemporal imaging of glutamate-induced biophotonic activities and transmission in neural circuits', *PLoS One*, 9, e85643.

Tegmark, M. (2000), 'Importance of quantum decoherence in brain processes', *Phys. Rev. E*, **61**, pp. 4194-4206.

Tegmark, M. (2017), *Life 3.0: Being Human in the Age of Artificial Intelligence* (New York: Penguin Random House).

Tononi, G. et al. (2016), 'Integrated information theory: from consciousness to its physical substrate', *Nature Reviews Neuroscience*, **17**, pp. 450-461.

Wade, P.D. *et al.* (1988), 'Mammalian cerebral cortical tissue responds to low-intensity visible light', *Proc. Natl. Acad. Sci. USA*, **85**, pp. 9322-9326.

Ward, L.M. (2011), 'The thalamic dynamic core theory of conscious experience', *Consciousness and Cognition*, **20**, pp. 464-486.

Warren, W.S. et al. (1998), 'MR Imaging contrast enhancement based on intermolecular zero quantum coherence', *Science*, **281**, pp. 247-251.

Watson, A.B. (1987), 'Efficiency of a model human image code', *J. Opt. Soc. Am. A*, **4**, pp. 2401-2417.

Wein, S. *et al*. (2018), 'Feasibility of efficient room temperature solid-state sources of indistinguishable single photons using ultra-small mode volume cavities', *Physical Review B*, **97**, 205418.

Wigner, E.P. (1967), 'Remarks on the mind-body question', in *Philosophical Reflections and Syntheses* (Berlin: Springer)

Yin, J. et al. (2017), 'Satellite-based entanglement distribution over 1200 kilometers', *Science*, **356**, pp. 1140-1144.

Zimmermann, M. (1989), 'The Nervous System in the Context of Information Theory', in Human Physiology, R.F. Schmidt and G. Thews (Berlin: Springer).